\documentclass[intlimits,twoside,a4paper]{article}

\usepackage[cp1251]{inputenc}

\usepackage[eqsecnum]{cmpj3}

\usepackage[version=4]{mhchem} 

\issue{2021}{24}{3}{33503}
\doinumber{10.5488/CMP.24.33503}
\title[Tetrahedrality, hydrogen bonding and the density anomaly of the central force water model. A Monte Carlo study]%
{Tetrahedrality, hydrogen bonding and the density anomaly of the central force water model. A Monte Carlo study 
\thanks{Dedicated to Prof.\ Yurii V.\ Kalyuzhny on the occasion of his 70$^\mathrm{th}$ birthday.}}
\author[V. Ravnik, B. Hribar Lee, O. Pizio, M. Luk\v{s}i\v{c}]{V. Ravnik\orcid{0000-0001-5605-3345
}\refaddr{label1},
        B. Hribar-Lee\orcid{0000-0002-9029-588X}\refaddr{label1}, O. Pizio\orcid{0000-0001-8333-4652}\refaddr{label2},  M. Luk\v{s}i\v{c}\orcid{0000-0001-7190-4013}\refaddr{label1}\thanks{Corresponding author: \email{Miha.Luksic@fkkt.uni-lj.si}.}}
\addresses{
\addr{label1} Faculty of Chemistry and Chemical Technology, University of Ljubljana, Ve\v{c}na pot 113, SI-1000 Ljubljana, Slovenia
\addr{label2} Instituto de Qu\'{i}mica, Universidad Nacional Autonoma de M\'{e}xico, Circuito Exterior, Ciudad Universitaria, 04510, Cd. de  M\'{e}xico, M\'{e}xico
}
\date{Received June 10, 2021, in final form July 22, 2021}

\begin{document}

\maketitle

\begin{abstract}
Monte Carlo computer simulations in the canonical and grand canonical statistical ensemble were used to explore the properties of the central force (CF1) water model. The intramolecular structure of the \ce{H2O} molecule is well reproduced by the model. Emphasis was made on hydrogen bonding, and on the tehrahedral, $q$, and translational, $\tau$, order parameters. An energetic definition of the hydrogen bond gives more consistent results for the average number of hydrogen bonds compared to the one-parameter distance criterion. At 300~K, an average value of 3.8 was obtained. The $q$ and $\tau$ metrics were used to elucidate the water-like anomalous behaviour of the CF1 model. The structural anomalies lead to the density anomaly, with a good agreement of the model's density with the experimental $\rho(T)$ trends. The chemical potential-density projection of the model's equation of state was explored. Vapour-liquid coexistence was observed at sufficiently low temperatures.      
\keywords{central force water model, hydrogen bonding, tetrahedraly, density anomaly, equation of state, Monte Carlo simulations}
%
%\pacs 05.10.Ln, 05.20.Jj, 05.70.Ce 
\end{abstract}

\section{Introduction}

Water is one of the most important substances on Earth. It is a key player in a vast variety of biological, geological, environmental, engineering, and technological processes. The majority of  interpretations of the experimentally observable properties of liquid water have been possible through modelling. Various models of water exist and can be broadly classified into three groups: quantum-chemical, atomistic, and coarse-grained~\cite{Brini_2017}. In 1975, Stillinger, Lemberg and Rahman proposed an isotropic coarse-grained central force (CF) model of water~\cite{Lemberg_1975, Stillinger_1975,Rahman_1975} and later introduced some improvements~\cite{Stillinger_1978}. In this model, water is regarded as a weak electrolyte, where oxygen and two hydrogens spontaneously form ion-triplets, i.e., a \ce{H2O} molecule. A collection of three pair potentials describing the hydrogen-hydrogen, hydrogen-oxygen, and oxygen-oxygen interactions are responsible for a non-linear geometry of \ce{H2O} molecule and its dipole moment. In a revised version of the model (CF1) by Haymet et al.~\cite{Duh_1995}, the set of potential parameters was adjusted in such a way as to bring the pressure of the model system at 25$^\circ$C closer to the atmospheric. The model has also been used to study solvation of ions~\cite{Bopp_1979,Jansco_1985,Holovko_2000,Druchok_2003} and of hydrophobic solutes~\cite{Arthur_1998,Arthur_1999} as well as at the interface with a planar wall (electrode)~\cite{Vossen_1994,Vossen_1995,Vossen_1997}.  

The most appealing feature of the central force models of water lies in the fact that a system of water molecules is treated as a mixture of partially charged particles representing oxygen and hydrogen atoms in the $1:2$ number ratio. Compared to commonly used atomistic water models (for example SPC and TIP models)~\cite{Brini_2017}, no angular dependent terms are included in the CF model. This is an advantage for simulations as well as for theoretical manipulations (for example, the Ornstein-Zernike equation can readily be applied to the model, cf. references~\cite{Lemberg_1976,Thuraisingham_1983,Ichiye_1988,Duh_1995}). Additionally, such a model can be used to  potentially address the autoprotolysis of water molecules (\ce{2H2O <=> H^+ + H3O^+}) and consequently the pH of the solution. 

Various CF models have been studied mainly by means of molecular dynamics (MD) computer simulations and integral equations. To our best knowledge, Monte Carlo computer simulations were used by Bresme~\cite{Bresma1998}, though not for the CF1 model proposed by Haymet et al.~\cite{Duh_1995}. In this work, the CF1 water model was examined by means of both canonical and grand canonical Monte Carlo simulations. The emphasis was made on the geometry of the water molecule,  hydrogen bonding, the tetrahedral and translational order parameters, the density anomaly and the equation of state.

\section{The model}

\begin{figure}[htb]
\centerline{\includegraphics[width=0.95\textwidth]{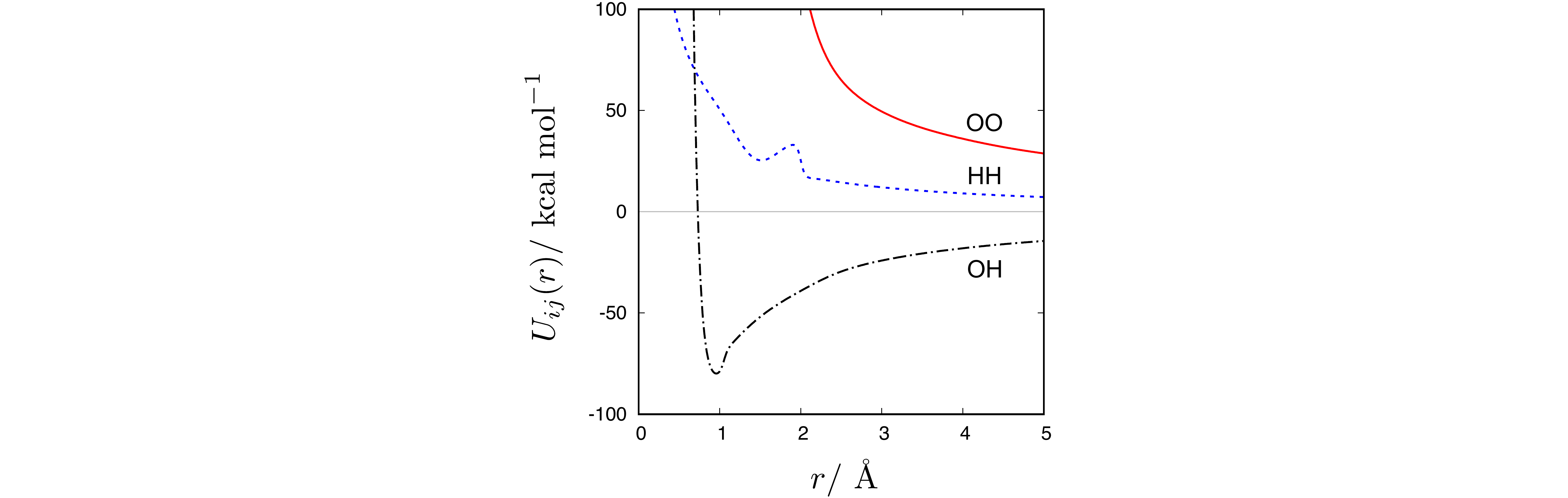}}
\caption{(Colour online) Distance dependence of the oxygen-oxygen (OO), hydrogen-hydrogen (HH), and oxygen-hydrogen (OH) pair potentials for the CF1 water model (cf.\ equations~(\ref{eq:OO}--\ref{eq:OH})).}
\label{fig:potential}
\end{figure}

In the CF1 model, the electroneutrality of \ce{H2O} molecule is ensured by assigning the partial charges of $q_\mathrm{H} = 0.32983e_0$ to hydrogens and $q_\mathrm{O} = -2q_\mathrm{H}$ to oxygen, where $e_0$ designates the elementary charge. The oxygen-oxygen ($U_\mathrm{OO}$), hydrogen-hydrogen ($U_\mathrm{HH}$), and oxygen-hydrogen ($U_\mathrm{OH}$) effective pair potentials are defined as~\cite{Duh_1995}:
\begin{align}
U_\mathrm{OO}(r) &= \frac{144.538}{r} + \frac{24082.38}{r^{8.8591}} - 0.25 \re^{-4(r-3.4)^2} - 0.25 \re^{-1.5(r-4.5)^2}, \label{eq:OO} \\[8pt]
U_\mathrm{HH}(r) &= \frac{36.1345}{r} + \frac{18}{1 + \re^{40(r-2)}} - 17 \re^{-7.62177(r-1.45251)^2}, \label{eq:HH} \\[8pt]
U_\mathrm{OH}(r) &= -\frac{72.269}{r} + \frac{6.23403}{r^{9.19912}} - \frac{10}{1 + \re^{40(r-1.05)}} - \frac{4}{1 + \re^{5.49305(r-2.2)}}, \label{eq:OH}
\end{align}
where $r$ is the particle-particle separation distance. Inserting the numerical value of $r$ in \AA ngstr\"{o}ms, the unit of the potential functions~(\ref{eq:OO}--\ref{eq:OH}) is kcal/mol (throughout the paper, energetic unit will be given in kJ/mol). All three potentials are displayed in figure~\ref{fig:potential}.

\subsection{Monte Carlo computer simulations}

The behaviour of the CF1 water model was explored using Monte Carlo computer simulations in the canonical $(N,V,T)$ and grand canonical $(\mu, V, T)$ statistical ensemble. $N$ represents the number of particles, $V$ the system's volume, $T$ the temperature, and $\mu$ the chemical potential.  All simulations were performed in a cubic simulation box with the box-edge length $L$, determined from the desired starting density, $\rho$, and number of water molecules $N$. The initial configuration of the system was obtained by a random insertion of particles (oxygens and hydrogens in number ratio 1:2). Periodic boundary conditions were implemented along with the Ewald summation method (Ewald parameter $\alpha$ was set to $5/L$)~\cite{Allen,FrenkelSmith}.

200 water molecules (i.e., 200 oxygens and 400 hydrogens) were used in canonical Monte Carlo~(CMC) simulations while the equilibrium number of water molecules in the grand canonical Monte Carlo~(GCMC) simulations was determined by the chemical potential, $\mu$. In the CMC simulations, the equilibration part involved $5 \cdot 10^8$ attempted particle displacement steps, followed by $1.5 \cdot 10^9$ steps for the production part. In a displacement step, a particle of the system (oxygen or hydrogen) was selected at random and moved to a new randomly selected position (max.\ displacement of 0.5~\AA). Standard Metropolis sampling algorithm was employed~\cite{Allen, FrenkelSmith}, with the acceptance probability for particle displacement being $\min \left [ 1, \exp \left ( -\beta \left [ U_\mathrm{new} - U_\mathrm{old} \right ] \right ) \right ] $. $\beta = 1/k_\mathrm{B} T$ ($k_\mathrm{B}$ is the Boltzmann constant), while $U_\mathrm{new}$ and $U_\mathrm{old}$ denote the energy of the system after and before the attempted particle move, respectively. In the GCMC simulations, a $1 \cdot 10^8$ steps long CMC simulation involving 200 water molecules was performed before the initiation of the grand canonical algorithm. The equilibration part in total involved $5 \cdot 10^8$ steps ($1 \cdot 10^8$ CMC and $4 \cdot 10^8$ GCMC), while production runs were $1.5 \cdot 10^9$ steps long. In the GCMC part, one attempted particle displacement was followed by an attempt to insert or remove a water molecule (one oxygen and two hydrogens). For the deletion of a water molecule, a random oxygen atom was  chosen alongside the two closest hydrogen atoms. These three particles were removed from the system. As for insertion, an oxygen atom was added to a random position inside the simulation box, then two hydrogens were added in a favourable geometry, so that the distance between the oxygen and hydrogen was in the interval $0.875 \leqslant r_\mathrm{OH} / \text{ \AA} \leqslant 1.075$ and the distance between the two inserted hydrogens was $1.3 \leqslant r_\mathrm{HH} / \text{ \AA}  \leqslant 1.7$. 
The acceptance probability for water insertion was $\min \left [ 1, Y_{N \rightarrow N+1} \right ]$ and for the deletion it was $\min \left [ 1, Y_{{N \rightarrow N+1}}^{-1} \right ]$, where~\cite{Torrie1908,Valleau1980}:
\begin{equation}
  Y_{N \rightarrow N+1} = \frac{N_{N}^\mathrm{O}! N_{N}^\mathrm{H}!}{N_{N+1}^\mathrm{O}! N_{N+1}^\mathrm{H}!} \exp \left [ B - \beta \left ( U_{N+1} - U_{N} \right ) \right ].
\end{equation}
Here, $N^\mathrm{O}_i$ and $N^\mathrm{H}_i$ denote the numbers of oxygens and hydrogens, respectively, in the state with less waters ($i = N$), i.e., before insertion or after removal, and the state with more particles ($i = N+1$), i.e., before removal or after insertion. Parameter $B$ is related to the chemical potential of \ce{H2O}, $\mu$, in the following way:
\begin{equation}
    B = \beta \mu  + 3 \ln \frac{L^3}{\Lambda_\mathrm{H}^2 \Lambda_\mathrm{O}},
\end{equation}
where $\Lambda_i = h/\sqrt{2\piup m_i k_\mathrm{B} T}$ is the thermal de Broglie wavelength for hydrogens ($i = \mathrm{H}$) or oxygens ($i = \mathrm{O}$). $h$ is the Planck constant and $m_i$ is the mass of the particle $i = \text{ H or O}$.    

The majority of simulations were performed at 300~K and water density of 1~g/mL. The temperature dependence of the studied quantities was explored up to 1000~K (at density of 1~g/mL), while density dependence at 300~K was studied in the range from 0.8 to 1.4~g/mL.

\section{Results and discussion}

In this section we present the results of Monte Carlo simulations of the CF1 water model. We studied the molecular geometry of the CF1 water, as well as the effect of temperature and density on it. We were also interested in hydrogen bonding. Two criteria were used to determine the average number of hydrogen bonds: the simplest one-parameter distance criterion based on the integration of the OH radial distribution function  and an energy criterion based on the pair energy distribution function. Alongside this, using the tetrahedral and translational order parameters we studied the tetrahedrality of the model water and the concomitant density anomaly. Using the GCMC simulations, the temperature dependence of the model water density was calculated and compared with experimental data. Furthermore, the $\mu - \rho$ projection of the equation of state for the CF1 water model was examined in a broad temperature range.

\subsection{Geometry of the CF1 water molecule}

\begin{figure}[htb!]
\centerline{\includegraphics[width=0.95\textwidth]{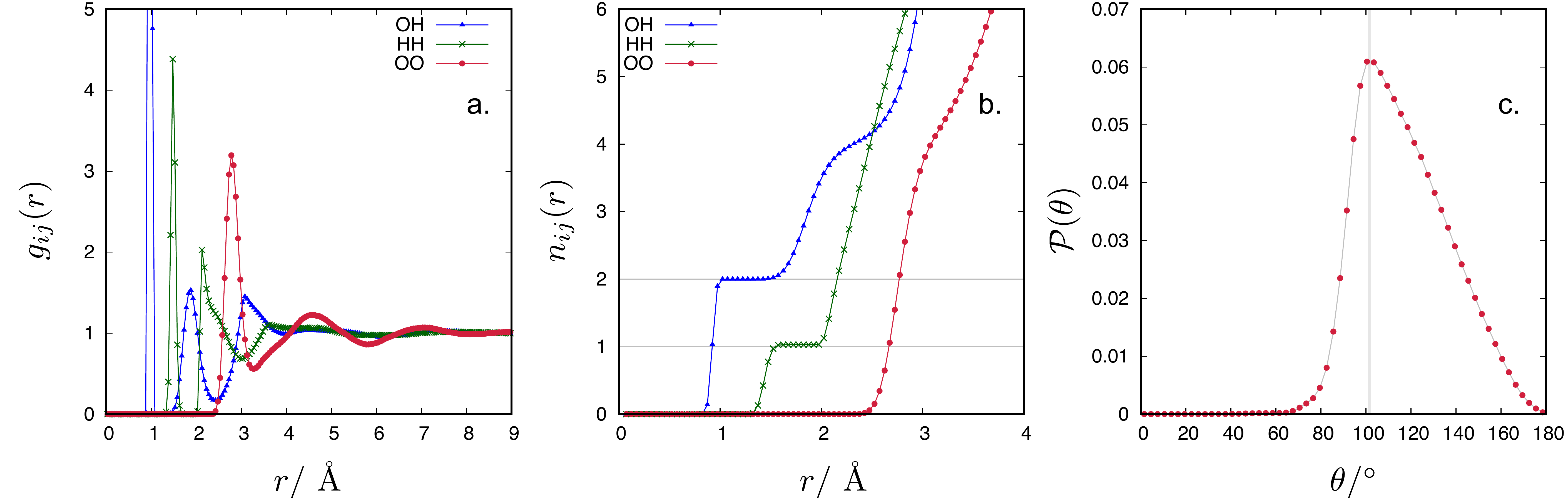}}
    \caption{(Colour online) The oxygen-hydrogen (blue triangles and lines), hydrogen-hydrogen (green stars and lines) and oxygen-oxygen (red circles and lines) radial distribution functions (panel a) and the corresponding running coordination numbers (panel b) for the CF1 water model with density 1~g/mL and temperature 300~K. The height of the first peak of OH RDF is $\sim 39$ (not shown). Panel c shows the unnormalized distribution of HOH bond angles, $\mathcal{P}(\theta)$. Symbols show the data obtained by canonical Monte Carlo simulations while lines are only guides to the eye.}
    \label{fig:structure}
\end{figure}

In figure~\ref{fig:structure}a we show the radial distribution functions (RDFs) for a CF1 water model at density 1~g/mL and temperature 300~K, while figure~\ref{fig:structure}b displays the running coordination numbers estimated from the corresponding RDFs, i.e. $n_{ij}(r) = 4\piup \rho_j \int_0^r g_{ij}(r')r'^2\mathrm{d}r'$, where $\rho_j$ is the number density of species $j$. The first peak in the oxygen-hydrogen RDF corresponds to the intramolecular O-H bond, while the first peak in the hydrogen-hydrogen RDF corresponds to the interaction between two hydrogens of a given water molecule. The intramolecular OH and HH coordination numbers are 2.0 and 1.0, respectively (indicated by a grey line in figure~\ref{fig:structure}b) in excellent agreement with the composition of the water molecule. The average O-H bond length is $l_\mathrm{OH} = 0.962$~\AA, and the average separation distance between the two hydrogens of the \ce{H2O} molecule is 1.496~\AA. From these two values, the HOH bond angle is $\theta = 102.1^\circ$. The $l_\mathrm{OH}$ is within the uncertainty interval of the experimentally determined O-H bond length for water in the liquid state, measured by neutron diffraction, ($0.970 \pm 0.005$)~\AA~\cite{Ichikawa1991}, and approximately 0.5\% larger than the experimental value of a free water molecule in the gas phase, ($0.9572 \pm 0.0003$)~\AA~\cite{Benedict1956}. The H-H distance is approximately 3.5\% shorter than the experimentally determined, $(1.55 \pm 0.01)$~\AA~\cite{Ichikawa1991}, which leads to smaller bond angle than observed in experiments, i.e., $106.1^\circ \pm 1.8^\circ$ for liquid water and $104.52^\circ \pm 0.05^\circ$ for water in the gas phase. From the un-normalized bond angle distribution function, $\mathcal{P}(\theta)$, shown in figure~\ref{fig:structure}c one can see that the reported bond angle of the CF1 water corresponds to the most probable value, although the distribution is quite broad and somewhat asymmetric towards larger $\theta$. In table~\ref{tab:WaterModelParameters}, the values of $l_\mathrm{OH}$ and $\theta$ for some atomistic water models frequently employed in computer simulations are given. The parameters of TIPxP ($x = 3,4,5$) water models are equal to experimental values for an isolated water molecule, while ST2 and SPC models assume a perfect tetrahedral angle and O-H bond length of 1~\AA. 

Experiments show that $l_\mathrm{OH}$ and $\theta$ remain unchanged within the experimental error in a broad temperature range (from 298~K to 473~K)~\cite{Ichikawa1991}. We simulated the CF1 model at 1~g/mL in the interval from 240~K to 500~K and observed no changes in $l_\mathrm{OH}$ and $\theta$ as well. The differences were less than 0.3\%. In addition, the intramolecular geometry also remained the same at $\rho = 0.9$ and $1.1$~g/mL in the reported temperature range.  

\begin{table}[htb]
\caption{Geometric parameters (O-H bond length and HOH bond angle) of some commonly used water models~\cite{Brini_2017}, and the experimental values for a water molecule in the gas~\cite{Benedict1956} and liquid phase~\cite{Ichikawa1991}.}
\label{tab:WaterModelParameters}
\vspace{2ex}
\begin{center}
\renewcommand{\arraystretch}{0}
\begin{tabular}{l|cc}
\hline\strut
Water (model)	        &		$l_\mathrm{OH}$, \AA	& $\theta$, $^\circ$	 \strut\\
\hline\strut
ST2  		          	&	1         	& 109.47		 \strut\\
SPC, SPC/E	            &	1         	& 109.47		 \strut\\
TIP3P, TIP4P, TIP5P 	&	0.9572    	& 104.52		 \strut\\
CF1                     &   0.962       & 102.1          \strut \\
\hline \strut
Experiment (gas)        &   0.9572      & 104.52      	 \strut\\
Experiment (liquid)     &   0.970       & 106.1     	 \strut\\
\hline   
\end{tabular}
\renewcommand{\arraystretch}{1}
\end{center}
\end{table}

\subsection{Hydrogen bonding in CF1 water}
The number of hydrogen bonds per water molecule can be approximately estimated from the OH coordination numbers determined at distances which correspond to the first and second minimum in the oxygen-hydrogen RDF (cf. figure~\ref{fig:structure}a). For~1 g/mL at 300~K, integrating the $g_\mathrm{OH}(r)$ up to the second minimum gives the coordination number of 4.14. Subtracting the value of the intramolecular OH coordination number (2.00) one obtains 2.14. The number of hydrogen bonds (donors and acceptors) between a test water molecule with its surrounding water molecules is therefore equal to $2 \times 2.14 = 4.28$. Experimental estimate for the number of hydrogen bonds per water molecules is approximately 3.5~\cite{Kumar2007}. Using the one-parameter distance  definition to determine the average number of hydrogen bonds of the CF1 water model leads to a larger number compared to experiment. In this work, we investigated the one-parameter distance criterion because such hydrogen bond definition is often used in analysing the results of integral equations. However, in simulations involving atomistic water models, commonly extended ``distance-only'' criteria are used where the O-O and O-H distance together with OH-O angle are taken into account as well~\cite{Kumar2007}. Various cut-offs for the hydrogen bond donor-acceptor distance and the hydrogen-donor-acceptor angle have been defined~\cite{Kumar2007}. Employing such a distance-angle definition for the hydrogen bond usually leads to a better agreement with experimental data.   

\begin{figure}[htb!]
\centerline{\includegraphics[width=0.95\textwidth]{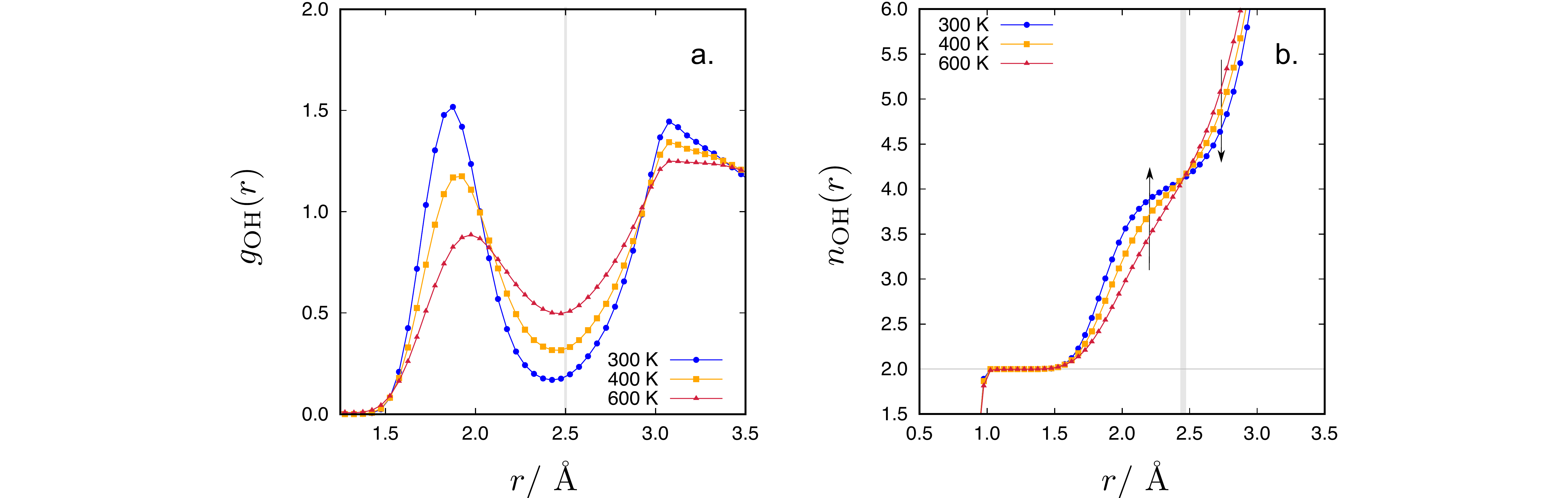}}
    \caption{(Colour online) The temperature dependence (300, 400, 600~K) of the intermolecular part of the oxygen-hydrogen radial distribution function (panel a), and the corresponding running OH coordination numbers (panel b). The density of the CF1 water model was 1~g/mL.}
    \label{fig:OHcoordnrs}
\end{figure}

In figure~\ref{fig:OHcoordnrs}, the temperature dependence of the intermolecular part of the $g_\mathrm{OH}(r)$ (panel a) and the running OH coordination number (panel b) are shown. We see that upon increasing the temperature of the system, the second minimum in the OH RDF slightly shifts towards larger $r$-values. The changes are, however, small: at 300~K, the minimum was found at 2.43~\AA, while at 400 and 600~K it was at 2.48~\AA.  Figure~\ref{fig:OHcoordnrs}b displays the sensitivity of the intermolecular OH coordination number on the $g_\mathrm{OH}(r)$ integration end-point, $r$. The running coordination numbers, $n_\mathrm{OH}(r)$, for different temperatures cross at $r =  (2.45 \pm 0.02$)~\AA. Very small differences in determining the integration end-point lead to opposing temperature trends in $n_\mathrm{OH}$ (indicated by arrows in figure~\ref{fig:OHcoordnrs}b). In table \ref{tab:Hbonds}, the average number of hydrogen bonds, determined by a distance criterion ($r = 2.5$~\AA), $\langle \text{H-bond} \rangle_r$, is given. Basically, no difference in the disruption of the hydrogen bonding network upon increasing temperature can be observed (all values of $\langle \text{H-bond} \rangle_r$ are approximately 4.3). Integration of the OH RDF is therefore not a very reliable way of determining the number of hydrogen bonds in the case of CF1 water model. It is also sensitive to the bin-width used in collecting the RDFs during the MC simulations (in our case, the bin-width was 0.05~\AA).

\begin{table}[htb]
\caption{The temperature dependence of the average number of hydrogen bonds estimated from the OH RDF (one-parameter distance criterion, 2.5~\AA), $\langle \text{H-bond}  \rangle_r$, and from the pair energy distribution function (with H-bond energy $-9.0$~kJ/mol), $\langle \text{H-bond} \rangle_E$, of the CF1 water model with density of 1~g/mL.}
\label{tab:Hbonds}
\vspace{2ex}
\begin{center}
\renewcommand{\arraystretch}{0}
\begin{tabular}{c|cccc}
  $T$ / K                  & 300  & 350  & 400 & 600 \strut\\
  \hline \strut
  $\langle \text{H-bond}  \rangle_r$    & 4.28 & 4.32 & 4.35 & 4.33 \strut\\
  $\langle \text{H-bond}  \rangle_E$    & 3.79 & 3.75 & 3.20 & 2.27 \strut\\
\end{tabular}
\renewcommand{\arraystretch}{1}
\end{center}
\end{table}

\begin{figure}[htb!]
\centerline{\includegraphics[width=0.95\textwidth]{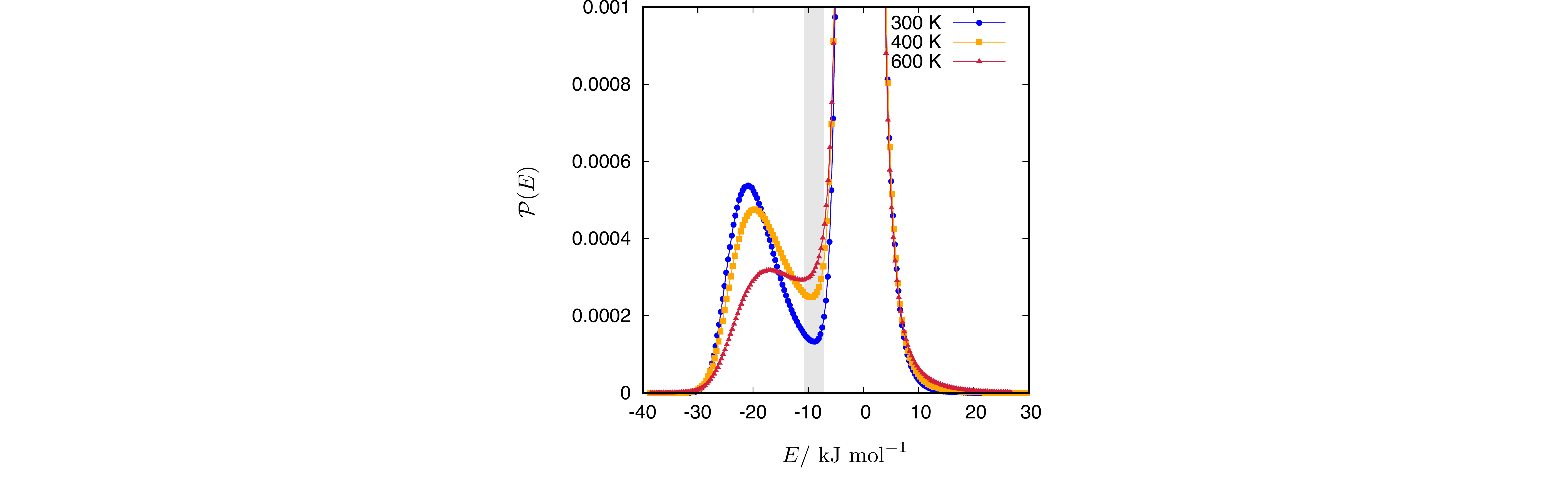}}
    \caption{(Colour online) The unnormalized pair energy distribution function, $\mathcal{P}(E)$, of the CF1 water model at 1~g/mL and 300, 400 and 600~K. }
    \label{fig:E2}
\end{figure}

We have therefore estimated the number of hydrogen bonds also from the energetic definition. In figure~\ref{fig:E2}, the pair energy distribution function, $\mathcal{P}(E)$, of the CF1 water model at various temperatures (300, 400, and 600 K) is given. The $\mathcal{P}(E)$ shows a minimum which fades-off with an increasing temperature. The minimum is located at $-8.7$, $-9.4$, and $-11.3$~kJ/mol for 300, 400, and 600~K, respectively. The values are somewhat less negative than for the atomistic water models. For example, TIP3P water model at 300~K has the pair energy minimum at approximately $-10.5$~kJ/mol~\cite{Jorgensen1983}. The minimum in the $\mathcal{P}(E)$ is a convenient energetic definition of a hydrogen bond~\cite{Kumar2007}. We have rather arbitrarily chosen that any pair of CF1 waters having energy at least $-9.0$~kJ/mol is considered to be hydrogen bonded. Table~\ref{tab:Hbonds} gives the temperature dependence of the average number of hydrogen bonds estimated from such an energetic criterion, $\langle \text{H-bond} \rangle_E$. The average number of hydrogen bonds at 300~K, determined in this way, is 3.79. This is much closer to the experimentally determined value for liquid water ($\sim 3.5$~\cite{Kumar2007}) than the value estimated from the $g_\mathrm{OH}(r)$. The increase of temperature also leads to the correct trend, namely the hydrogen-bonding network weakens with increasing temperature (average number of hydrogen bonds decreases with temperature). From the results presented in this subsection, we conclude that in case of the CF1 water model, the pair energy criterion of the average number of hydrogen bonds gives more consistent results compared to the distance criterion. 

\subsection{Tetrahedrality of the CF1 water}

\begin{figure}[htb!]
\centerline{\includegraphics[width=0.95\textwidth]{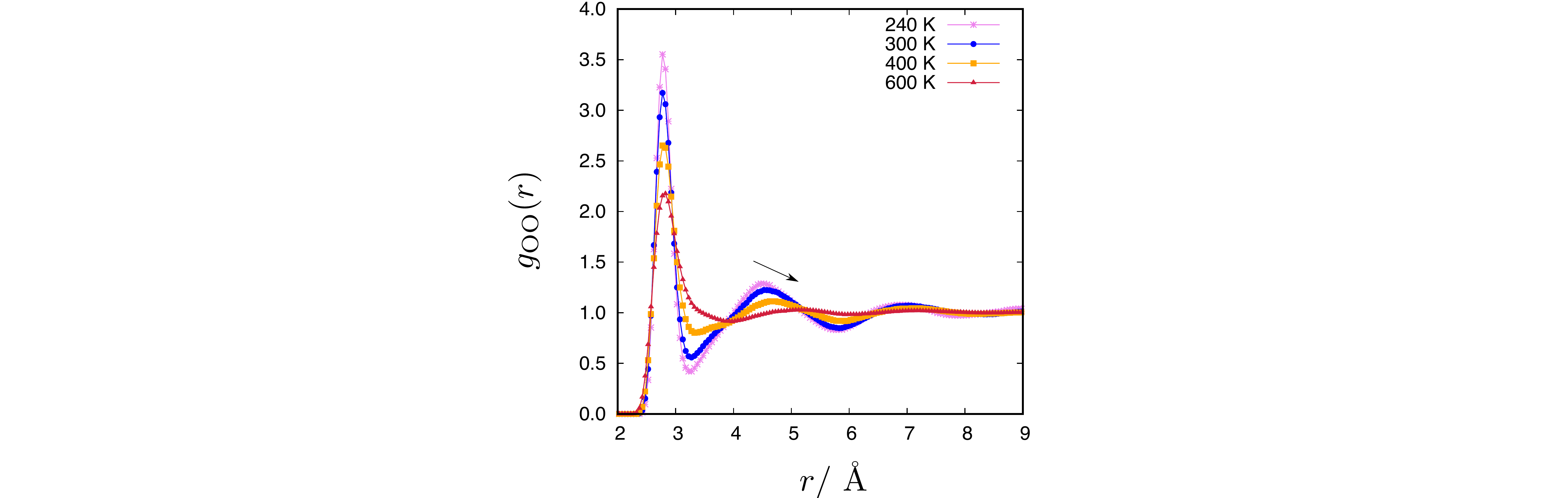}}
    \caption{(Colour online) The temperature dependence of the oxygen-oxygen pair distribution function, $g_\mathrm{OO}(r)$, of the CF1 water model at 1~g/mL.}
    \label{fig:grOO}
\end{figure}

Water is a tetrahedral liquid~\cite{Jabes2012,Brini_2017,Urbic2018}.The first nearest neighbour water molecules correspond to the first peak in the $g_\mathrm{OO}(r)$. The second peak in the oxygen-oxygen RDF, however, is argued to indicate the presence of tetrahedral order of waters in the liquid state~\cite{Soper1986,Duh_1995}. Figure~\ref{fig:grOO} shows how the $g_\mathrm{OO}(r)$ changes with temperature. By increasing the temperature the height of the first and second peaks diminishes. The location of the second maximum slightly shifts towards larger distances (indicated by the arrow) and becomes only marginally pronounced at high temperatures (600~K). This indicates that the tetrahedral order of CF1 water diminishes with an increasing temperature. As also seen  from the average number of hydrogen bonds, $\langle \text{H-bond} \rangle_E$ (table~\ref{tab:Hbonds}), the hydrogen bonding network in the CF1 water model weakens with temperature.    

A variety of different order parameters have been proposed to characterise the local structure of water in the liquid state. Here, we explore the temperature and density dependence of the so-called angular tetrahedral order parameter, defined as~\cite{Chau1998,Errington2001}:
\begin{equation}
    q = 1 - \frac{3}{8} \sum_{i = 1}^3 \sum_{j = i+1}^4 \left ( \cos \psi_{ij} + \frac{1}{3} \right )^2,
\end{equation}
where $\psi_{ij}$ denotes the angle between the vectors joining the oxygen atom of a selected test water and its nearest neighbour oxygen atoms $i$ and $j$. It describes the angular ordering of the first hydration shell waters. For an ideal gas, the average value of $q$ equals zero, while for a perfect tetrahedral arrangement it equals one. High values of $q$ can be found for an angularly ordered but radially disordered hydration shell as well as in cases where there is not a clear separation between the first and second shell. By cooling the system, it becomes more tetrahedral, although it still differs from the ice-like configuration.

\begin{figure}[htb!]
\centerline{\includegraphics[width=0.95\textwidth]{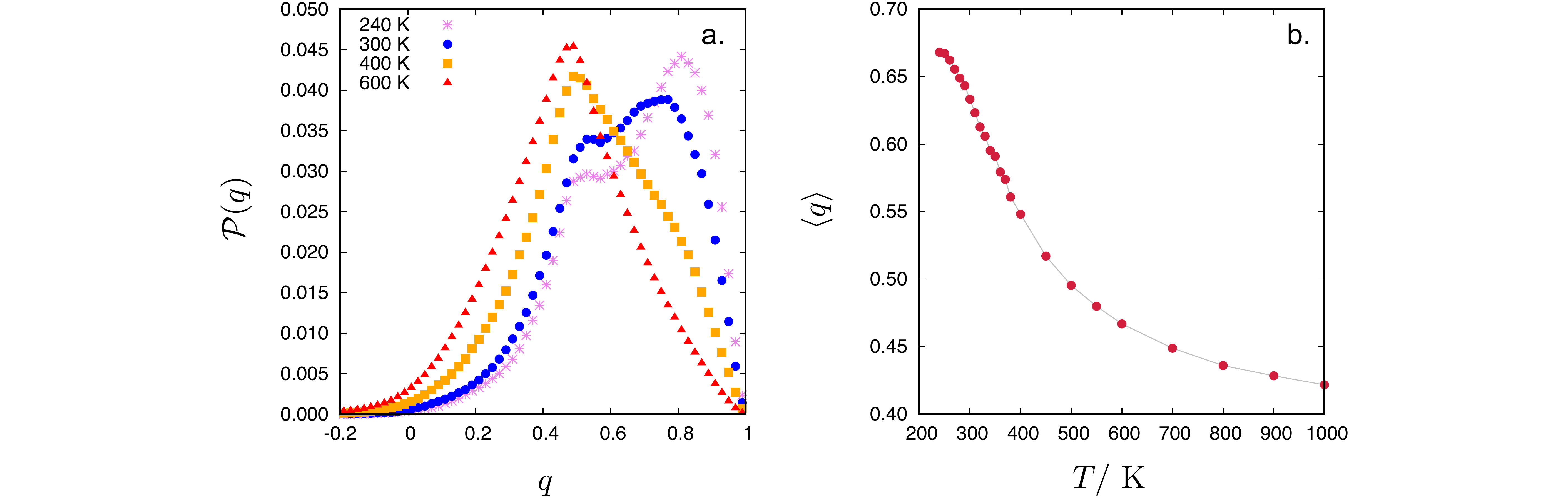}}
    \caption{(Colour online) Distribution of the tetrahedral order parameter, $\mathcal{P}(q)$, for the CF1 water model at 1~g/mL for four different temperatures (240, 300, 400 and 600~K) (panel a) and the temperature dependence of the average tetrahedral order parameter, $\langle q \rangle$, at 1~g/mL (panel b).}
    \label{fig:qtedistribution}
\end{figure}

In figure~\ref{fig:qtedistribution}a, we show the distribution of the tetrahedral order parameter, $\mathcal{P}(q)$, of the CF1 water model at 1~g/mL for four different temperatures (240, 300, 400 and 600~K). At low temperatures (240 and 300~K), a shoulder-like distribution is observed. Such a shape is characteristic of cases where compatible fractions of two distinct local environments exist: a tetrahedral and nontetrahedral~\cite{Errington2001}. The most probable value of $q$ [indicated by the maximum on the $\mathcal{P}(q)$] is 0.81 at 240~K and moves to 0.77 by increasing the temperature to 300~K. A bi-modal distribution observed at 240 and 300~K does not mean that in the liquid there exist two types of arrangements that would be energetically favoured. It only means that the arrangement of four nearest neighbours of the central water molecule can be predominantly ice-like structured (the peak at high $q$ values) or unstructured (smaller peak at lower $q$ values)~\cite{Shoper2000,Errington2001}. Increasing the temperature even further (400 and 600~K) leads to the loss of strong tetrahedrality and at 600 K a completely uni-modal distribution is observed with the most probable $q$ value of 0.49. In figure~\ref{fig:qtedistribution}b, the temperature dependence of the average tetrahedral order parameter, $\langle q \rangle$, is shown. The $\langle q \rangle$ monotonously decreases with an increasing temperature, in line with temperature trends in $g_\mathrm{OO}(r)$. It is worth noting that at 240~K the system is still in the liquid state rather than a solid ice. 

\subsection{Density anomaly of the CF1 water model}

For all liquids, the local order is determined by the pair correlations. To evaluate the degree of ordering due to pair correlations between oxygen atoms, the so-called translational order parameter was calculated as~\cite{Kumar22130,Truskett2000}:
\begin{equation}
    \tau = \frac{1}{r_\mathrm{c}}  \int_{0}^{r_\mathrm{c}} | g_\mathrm{OO}(r) - 1 | \mathrm{d} r,
\end{equation}
where $g_\mathrm{OO}(r)$ denotes the oxygen-oxygen pair correlation function, and $r_\mathrm{c}$ is an appropriate cut-off distance. In our case, $r_\mathrm{c}$ was chosen as a half of the simulation box edge-length. For an ideal gas, $\tau$ is zero. $\tau$ increases as the system becomes more ordered. The degree of correlation between the average tetrahedral order parameter, $\langle q \rangle$, and the average translational order parameter, $\langle \tau \rangle$, can be used to explore the structural anomalous region in the density-temperature plane. This region approximately encloses the region of the thermodynamic (density) anomaly~\cite{Errington2001}.

\begin{figure}[htb!]
\centerline{\includegraphics[width=0.95\textwidth]{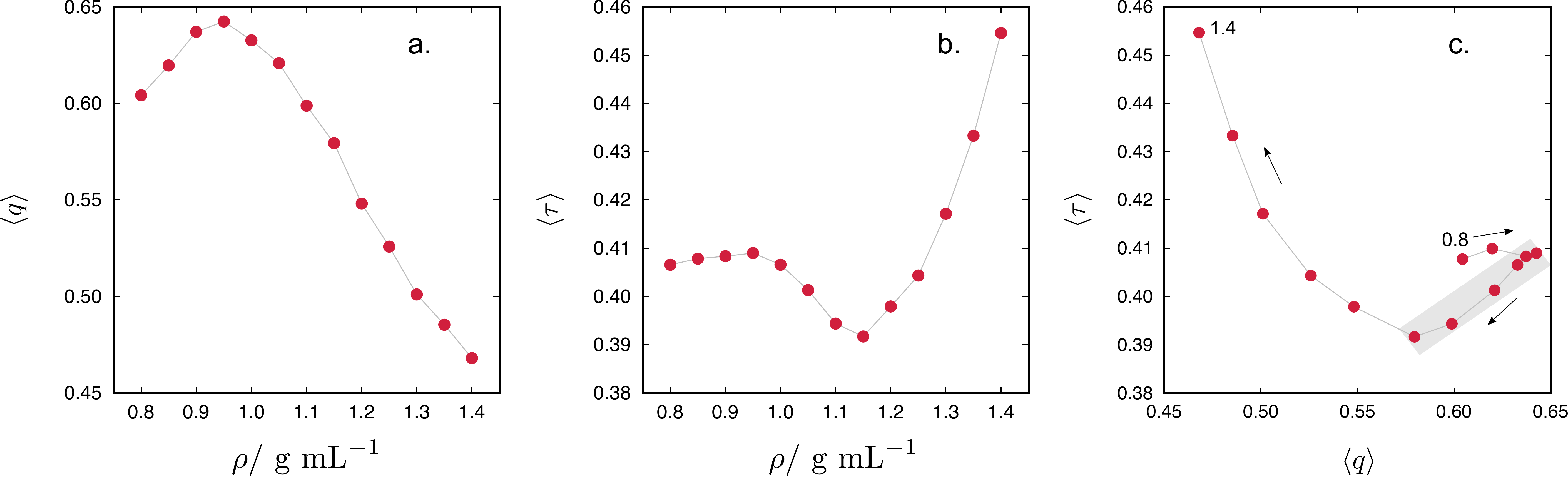}}
    \caption{(Colour online) Density dependence of (a) average tetrahedral order parameter, $\langle q \rangle$, and (b) average translational order parameter, $\langle \tau \rangle$. Correlation of $\langle q \rangle$ with $\langle \tau \rangle$ is given in panel (c). Arrows indicate increasing density (from 0.8 to 1.4~g/mL). Data apply for $T = 300$~K.}
    \label{fig:qtettau}
\end{figure}

We have performed canonical Monte Carlo simulations of the CF1 water model at 300~K for different densities, ranging from 0.8 to 1.4~g/mL. The dependence of  $\langle q \rangle$ and  $\langle \tau \rangle$ on the density is given in figures~\ref{fig:qtettau}a and~\ref{fig:qtettau}b, respectively. The shape of $\langle q \rangle$ on $\rho$ is characteristic of tetrahedral liquids: it exhibits a maximum at densities approaching the density of the ice-like structures. In our case, the maximum is located at 0.95~g/mL. The translational order parameter has a shallow maximum at low densities followed by a deeper minimum at 1.15~g/mL. The maximum in $\langle \tau \rangle$ at 0.95~g/mL coincides with the maximum in $\langle q \rangle$. The locus of maxima and minima in the density dependence of $\langle \tau \rangle$ is characteristic of the structurally anomalous region~\cite{Agarwal2011}. In addition, such a region also implies a linear correlation between the $\langle q \rangle$ and $\langle \tau \rangle$~\cite{Errington2001,Agarwal2011}, as seen for the CF1 water model in figure~\ref{fig:qtettau}c (the anomalous region is marked with a grey rectangle). 

\begin{figure}[htb!]
\centerline{\includegraphics[width=0.95\textwidth]{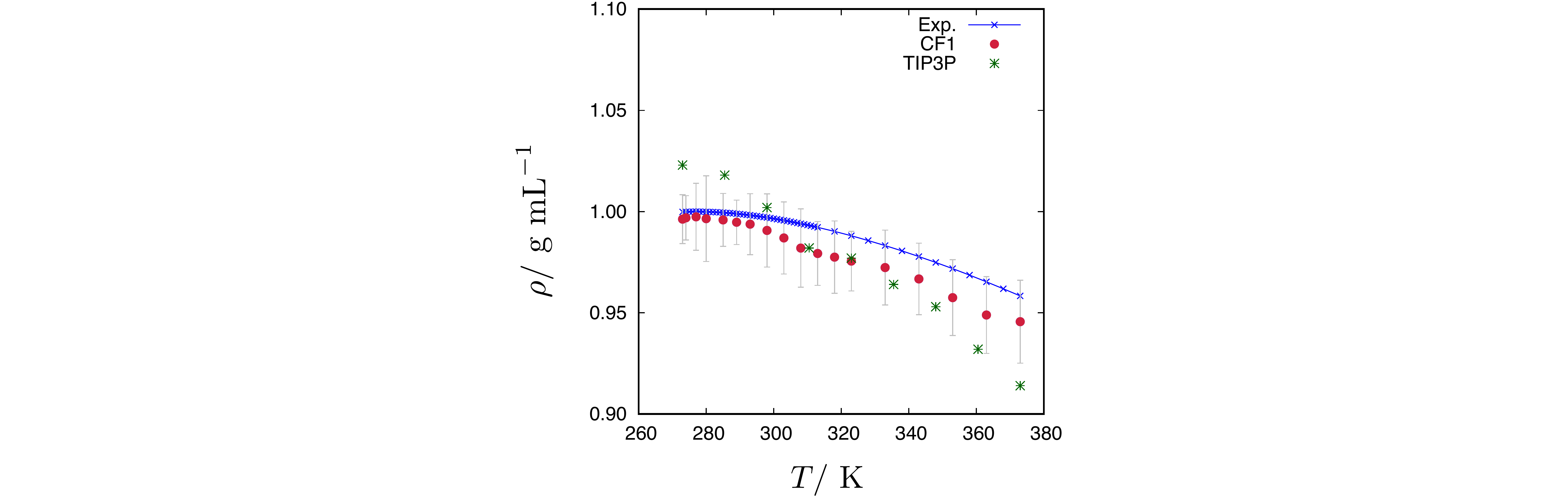}}
    \caption{(Colour online) Temperature dependence of the water density. Displayed are experimental data for liquid water (blue x's), CF1 water model (red circles) and TIP3P water model (green stars). For the CF1 model, the results were obtained by GCMC simulations using a linear dependence of the chemical potential on temperature. Data for TIP3P and experiment was taken from~\cite{Jorgensen1998}.}
    \label{fig:denstemp}
\end{figure}

Next, we explore how the CF1 water model captures the correlation between the local structural order and density anomaly. It is well known that liquid water at ambient pressure exhibits a density anomaly, i.e., the density reaches a maximum value at around 277~K~\cite{Brini_2017}. The origin of the anomalous behaviour lies in increased structural fluctuations upon cooling water down to the Widom line~\cite{Nilsson2015}. We used GCMC simulations to explore the temperature dependence of CF1 model density in the range from 273 to 373~K. A linear approximation for the dependence of the chemical potential, $\mu$, on temperature~\cite{Job2006} was assumed in the whole temperature range. In figure~\ref{fig:denstemp}, we compare our calculations with the experimental data for liquid water. Calculated density at a given temperature, $T$, corresponds to the following chemical potential: $\mu = -589.0 \text{ kJ/mol} - 0.08 \text{ kJ/mol\,K} \cdot T$. We see that a linear approximation for $\mu(T)$ gives good agreement between the calculated and experimental density of water. The difference between the calculated and experimental density is only 0.25\% at 277~K, 0.65\% at 298~K, and around 1.3\% at 373~K. These differences are smaller than the uncertainties in Monte Carlo calculations, which are of the order of 1.5--2\%. It is also worth mentioning that in the entire temperature range studied, the relative differences in experimental densities are small. For example, experimental densities for water at 277 and 373~K differ by approximately only 4.2\%. The CF1 water model also exhibits a small density maximum that can be viewed as the density anomaly (true density anomaly, however, should be explored at constant pressure conditions). Frequently used atomistic models of water (for example SPC or TIP3P) do not give such good results (data for TIP3P water model are given in figure~\ref{fig:denstemp}).

\subsection{Preliminary insights into the chemical potential-density projection of the equation of state for the CF1 water model}

Here, we also present  some preliminary insights into the behaviour of the $\mu - \rho$ projection of the equation of state of the CF1 water model in a limited temperature interval. A more extensive investigation of the model's equation of state --- namely exploration of the phase transition, determination of the coexisting densities, construction of the pressure-temperature projection of the phase diagram --- calls for a separate study.

\begin{figure}[htb!]
\centerline{\includegraphics[width=0.95\textwidth]{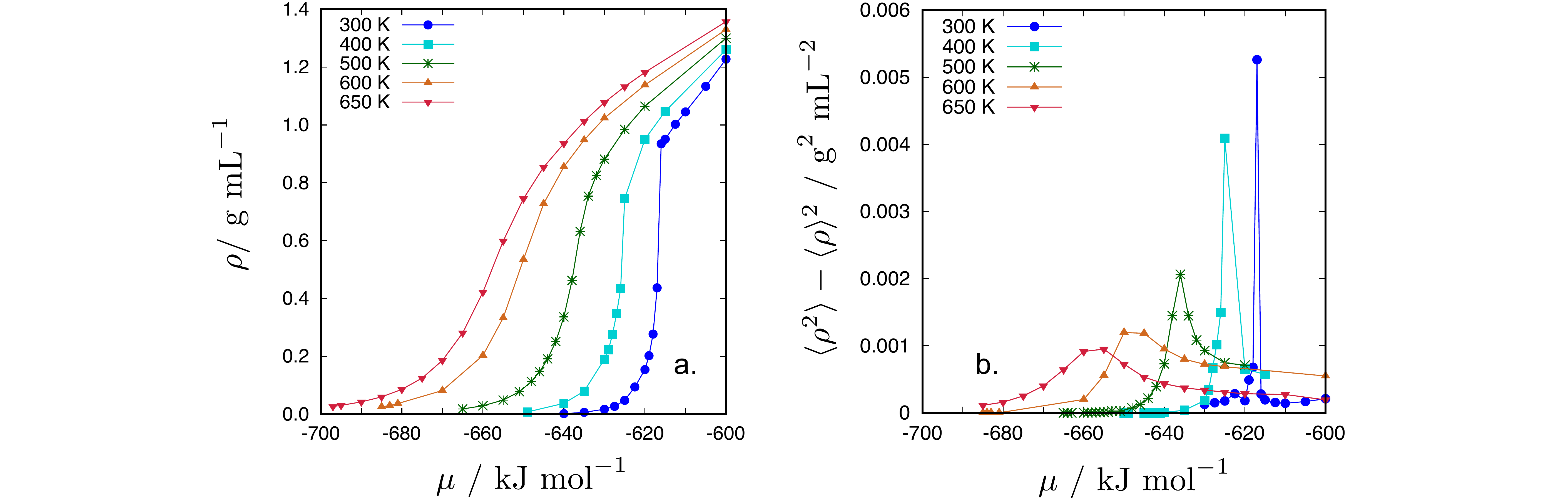}}
    \caption{(Colour online) (a) The $\mu-\rho$ projection of the equation of state of the CF1 water model, and (b) the density fluctuations, $\langle \rho^2 \rangle - \langle \rho \rangle^2$, at various temperatures.} 
    \label{fig:phase}
\end{figure}

Employing the GCMC simulations, we determined the dependence of the system's density, $\rho$, on the chemical potential, $\mu$, at different values of temperature (ranging from 300 to 650~K). Results are shown in figure~\ref{fig:phase}a: at sufficiently high temperatures (above 400~K), the dependence of the $\rho$ on $\mu$ is monotonous, i.e., the density increases by increasing the chemical potential. The $\rho(\mu)$ is a single-valued sigmoid-like function (a given value of the chemical potential determines a unique density of the system). Such a behaviour is characteristic of a system above the vapour-liquid critical point. By decreasing the temperature, the change in density (from gas-like to fluid-like) becomes more steep, until two distinct branches occur: one corresponds to the gas phase and the other to the liquid phase. For a given value of the chemical potential, two densities (low and high) can be found, and this value of $\mu$ corresponds to the liquid-vapour coexistence. Hysteresis loop upon approaching the phase transition from the gas phase densities is observed. 

The fluctuations in the system's density, $\langle \rho^2 \rangle - \langle \rho \rangle^2$, as a function of the chemical potential, $\mu$, are shown for different temperatures in figure~\ref{fig:phase}b. We can see that by decreasing the temperature (from 650 down to 300~K), the function starts to exhibit sharper peaks.  The increasing fluctuations in density are an indication that by further decreasing the temperature, the system will undergo the phase transition (in this case the vapour-liquid phase transition). The chemical potential at which the function diverges corresponds to the phase transition. At this point, we need to stress that our system was rather small, composed of approximately only 200 water molecules. This fact prohibits one to adequately address particle fluctuations.  

The standard GCMC simulations employed in this work along with the small size of the system do not allow for a precise determination of the critical point. Further temperature examination of the $\rho(\mu)$ dependence for a system with a larger number of water molecules is needed as well as the use of more sophisticated sampling techniques (for example~\cite{Rzysko2010}) to construct the density-temperature projection of the phase diagram. The only report on the phase diagram of a modified version of the central force model used in this work has been reported in~\cite{Bresme2001}. The author used nonequilibrium molecular dynamics simulation to calculate the vapour-liquid coexistence by explicitly simulating the vapour-liquid interface. The initial configuration of 540 water molecules arranged in a face-centered cubic lattice was subjected to a heat flux along one of the simulation box directions, leading to the generation of the liquid and vapour phases. To obtain the equilibrium phases, the temperature gradient was removed and constant temperature along the box was maintained. From local densities in different layers, the coexisting densities were determined. Good agreement of the vapour-liquid coexistence curve with experimental data for water was observed. Scaling law with the Ising exponent was used to obtain the critical temperature ($T_\mathrm{c} = 630.4$~K) and critical density ($\rho_\mathrm{c} = 0.337$~g/mL) of the model, and critical pressure ($p_\mathrm{c} = 337$~bar) was determined by using the Clausius-Clapeyron equation. From our calculations, we can only conclude that the critical temperature of the employed CF1 water model is quite below the experimental value for water (647.1~K). The reasons need to be addressed in the future: it can be due to a small size of the system used in our simulations, due to the used GCMC methodology, or perhaps due to differences in the model (cf.\ figure 1 of~\cite{Bresme2001} for comparison of potentials used in this work and by Bresme).

\section{Conclusions}

We have studied the CF1 model of water using the canonical and grand canonical Monte Carlo computer simulations. We have shown that the CF1 model reproduces well the molecular geometry of the water molecule. Our focus was on the hydrogen bonding of the model water molecules and on the temperature trends in the average number of hydrogen bonds. By using the energetic criterion (i.e., the minimum in the pair energy distribution function) instead of the commonly used one-parameter distance criterion (integration of the OH radial distribution function), we obtained a better agreement of the average number of hydrogen bonds with experiment as well as correct temperature trends. We also studied the evolution of the tetrahedral and translational order parameters as a function of water density. Structural anomalies were observed at 300~K leading to correct temperature trends in the density of water (so-called density anomaly). The $\mu - \rho$ projection of the equation of state allowed us to address the vapour-liquid phase transition. 

In the future, a more detailed analysis of hydrogen bonding in the CF1 water model is needed. This includes an extension of the one-parameter distance criterion to distance-angle criteria, which are often used in explicit water computer simulations. In addition, knowledge of the $\mu - \rho$ projection of the equation of state in a wide range of phase space is extremely important, for example, to study adsorption phenomena. Our results of the equation of state of the model are only preliminary. Simulations of systems with a much larger number of CF1 water molecules will be required, together with the use of more sophisticated simulation techniques. This will allow the determination of the coexisting densities and the construction of the $p-T$ phase diagram. Comparison with the experimental phase diagram of water will provide opportunities to modify the model's potential functions to achieve a better agreement of the model with real water.

\section*{Acknowledgements}

M.\ L.\ and B.\ H.-L.\ acknowledge the financial support from the Slovenian Research Agency (research core funding No.\ P1-0201) and the National Institutes of Health RM1 award ``Solvation modelling for next-gen biomolecule simulations'' (grant No. RM1GM135136). 

%% Type in your references using {thebibliography} environment 
%% or create them from your bibtex database using cmpj.bst style.

	\ukrainianpart

\title{Тетраедричність, водневі зв'язки та аномалії густини у моделі води з центральними взаємодіями. Дослідження методом Монте-Карло}
\author[В. Равнік, Б.Грібар-Лі, О. Пізіо, М. Лукшич]{В. Равнік\refaddr{label1},
	Б. Грібар-Лі\refaddr{label1}, О. Пізіо\refaddr{label2}  М. Лукшич\refaddr{label1}}
\addresses{
	\addr{label1} Факультет хімії та хімічних технологій університету Любляни, вул. Вечна 113, SI-1000, Любляна, Словенія
	\addr{label2} Інститут хімії Національного автономного університету Мехіко, Сіркіто Екстеріор, Сюідад Університаріа, 04510, Мехіко, Мексика}

\makeukrtitle

\begin{abstract}
	Моделювання Монте Карло у канонічному та великому канонічному ансамблях застосовується для дослідження властивостей води у моделі з центральними взаємодіями (ЦВ1). Внутрішня структура молекули \ce{H2O} добре відтворюється даною моделлю. Основна увага зосереджується на водневих зв'язках, та параметрах порядку тетраедричності $q$ і трансляційному  $\tau$. Визначення водневого зв'язку в енергетичних термінах дає кращі результати для середньої кількості цих зв'язків у порівнянні з критерієм однопараметричної відстані. При 300~K було отримано середнє значення 3.8. Метрики $q$ і $\tau$ використовувались для висвітлення водоподібної аномальної поведінки моделі ЦВ1. Структурні аномалії ведуть до аномалії густини, причому густина моделі добре узгоджується з експериментальними тенденціями для $\rho(T)$. Було досліджено хімічну складову залежності ``потенціал-густина'' у рівнянні стану даної моделі.	Співіснування пари з водою спостерігається при достатньо низьких температурах.

\keywords{модель води з центральними взаємодіями, водневі зв'язки, тетраедричність, аномалії густини, рівняння стану, моделювання Монте Карло}
\end{abstract}

\lastpage
\end{document}